\documentclass[twocolumn,caps,showpacs]{revtex4}

\usepackage{graphicx,color}
\usepackage{latexsym}
\usepackage{amsmath,amssymb}        
\usepackage[draft=false]{hyperref}
\usepackage{mathrsfs}
\DeclareSymbolFontAlphabet{\mathrsfs}{rsfs}
\DeclareMathAlphabet{\mathcal}{OMS}{cmsy}{m}{n}

\begin{document}


\title{Accretion of Supersonic Winds on Boson Stars}


\author{M. Gracia-Linares and F. S. Guzm\'an}
\affiliation{Laboratorio de Inteligencia Artificial y Superc\'omputo. 
		Instituto de F\'{\i}sica y Matem\'{a}ticas, Universidad
              Michoacana de San Nicol\'as de Hidalgo. Morelia, Michoac\'{a}n,
              M\'{e}xico.}


\date{\today}


\begin{abstract}
We present the evolution of a supersonic wind interacting with a Boson Star (BS) and compare the resulting wind density profile with that of the shock cone formed when the wind is accreted by a non-rotating Black Hole (BH) of the same mass. The physical differences between these accretors are that a BS, unlike a BH has no horizon, it does not have a mechanical surface either and thus the wind is expected to trespass the BS. Despite these conditions, on the BS space-time the gas achieves a stationary flux with the gas accumulating in a high density elongated structure comparable to the shock cone formed behind a BH. The highest density resides in the center of the BS whereas in the case of the BH it is found on the downstream part of the BH near the event horizon. The maximum density of the gas is smaller in the BS than in the BH case. Our results indicate that the highest density of the wind is more similar on the BS to that on the BH when the BS has high self-interaction, when it is more compact and when the wind velocity is higher. We expect this and similar analyses help to know whether BSs can still be Black Hole mimickers or can be ruled out.
\end{abstract}





\maketitle

\section{Introduction}
\label{sec:introduction}

Boson Stars (BSs) are solutions of Einstein's equations sourced by a complex scalar field. Originally these solutions were spherically symmetric, stationary and considering a free field scalar field potential \cite{Kaup1968,RuffiniBonazzola1969}. Later on these systems were studied with a self-interaction term \cite{MColpi}. These systems were evolved and their stability in full non-linear general relativity was confirmed to be consistent with perturbation theory \cite{SeidelSuen1990}, including the case with self-interaction \cite{Balakrishna1998}. Under more refined conditions, the critical collapse of BSs near the onset of collapse was studied \cite{scott}. At some point BSs worked as special test cases for numerical relativity codes \cite{Dramlisch,Guzman2004} and eventually their role as potential Gravitational Wave (GW) sources was studied \cite{BalakrishnaGuzman}, including the Binary Boson Star system \cite{BBS}. There are various reviews on Boson Stars describing in detail their history focused on stationary solutions \cite{Jetzer1992,topical-review}, and a recent review that includes a more dynamical and numerical relativity oriented analysis \cite{PalenzuelaSteve}.

The closest role of Boson Stars to astrophysical scenarios has been that of Black Hole mimickers, which consists in analyzing the possibility that BSs could play the role of Black Holes due to their compactness, transparency to light and that they only interact gravitationally with ordinary matter. After the recent discovery of Gravitational Waves \cite{GW150914,GW151226}, and the high likelihood that the source  is a Binary Black Hole system, BSs have little room as Black Hole mimickers in the strong field regime, although there is still room for other compact objects to source these GWs \cite{ref1,ref2}. Nevertheless, BSs could maintain an accretion disk whose spectrum could be that of a Black Hole of the same mass \cite{diego-acc,Guzman2006}. Restrictions of BSs as mimickers involve the analysis of the differences in the way they deflect light as compared to deflection due to Black Holes \cite{RuedaGuzman2009} and the different GW signals of perturbed BSs \cite{BalakrishnaGuzman}. These two restrictions are to be refined under high resolution observations with for instance the Event Horizon Telescope \cite{EHTelescope}. Also, with the discovery of GW150914 and GW151226, the  ringdown modes have been shown to be observable by the LIGO detectors and could be comparable with those of Boson Stars and possibly their corresponding tails \cite{BSQNMs}.
Still, Boson Stars in the weak field regime have an important chance in the context of dark matter, not only as gigantic boson star halos \cite{BECHAlos}, but also as axion stars playing the role of mini-MACHOS for physically acceptable boson masses \cite{Chavanis}.

Meanwhile, in this paper we present another potential difference between Black Holes and Boson Stars in a Bondi-Hoyle accretion process. It is known that when a supersonic wind approaches a black hole space-time, a high density shock cone is formed in the downstream zone behind the black hole, a process that has been studied in rotating black hole space-times with slab symmetry (e.g. \cite{BHRezzolla, BHAlexLoraGuzman}), where the vibrations and oscillations of the shock cone show a potential to play the role of QPOs \cite{BHRezzolla}, a possibility that has also been explored for winds approaching the black hole through one of its poles \cite{BHFabio}. More recently, the evolution of the shock cone has been studied for slow winds in full 3D independently of the orientation of the wind with respect to the axis of rotation of the black hole \cite{Mycol2015}. Other scenarios include more elaborated wind configurations, for instance with velocity gradients \cite{CruzLora2016}.

Specifically, we explore the formation of a high density structure of the gas around a Boson Star and compare it with the shock cone formed behind a black hole of the same mass. For this we solve the relativistic Euler equations numerically under the conditions that the wind does not distort the background space-time of a BS or a BH and that it obeys an ideal gas equation of state. We analyze two regimes of velocity, a slow supersonic wind or equivalently small accretor case and a fast supersonic wind that corresponds to a big accretor size \cite{foglizzo2005}.

The paper is organized as follows. In Section \ref{sec:system} we present a brief description of Boson Stars, the selection of BSs for our analysis and the initial conditions of the wind. In Section \ref{sec:results} we present the results of the Bondi-Hoyle accretion onto BSs and compare it with the results of the shock cone on a BH. Finally in Section \ref{sec:conclusions} we draw some conclusions.

\section{Description of the system}
\label{sec:system}

\subsection{Boson Star space-times}
\label{subsec:BSs}

BSs are constructed from the Lagrangian density of a complex scalar field minimally coupled to gravity ${\cal L} = -\frac{R}{16\pi G} + g^{\mu \nu}\partial_{\mu} \phi^{*} \partial_{\nu}\phi + V(|\phi|^2)$, where $\phi$ is the scalar field and $V$ its potential of 
self-interaction \cite{Jetzer1992,topical-review}. The resulting stress-energy tensor is
$T_{\mu \nu} = \frac{1}{2}[\partial_{\mu} \phi^{*} \partial_{\nu}\phi +
\partial_{\mu} \phi \partial_{\nu}\phi^{*}] -\frac{1}{2}g_{\mu \nu}
[\phi^{*,\alpha} \phi_{,\alpha} + V(|\phi|^2)]$. BSs are related to the potential $V=m^2 |\phi|^2 + \frac{\lambda}{2}|\phi|^4$, although the name Boson Star has been applied to solutions using other types of potentials (see e.g. \cite{diego-f}). The mass of the boson is $m$ and $\lambda$ is the coefficient of self-interaction of the field. The evolution equation for the scalar field is the Klein-Gordon equation 

\begin{equation}
\left(
\Box - \frac{dV}{d|\phi|^2}
\right)\phi = 0,
\end{equation}

\noindent where $\Box \phi=\frac{1}{\sqrt{-g}}\partial_{\mu}[\sqrt{-g}
g^{\mu\nu}\partial_{\nu}\phi]$. BSs are spherically symmetric solutions of Einstein's equations for the above stress energy tensor and the Klein-Gordon equation, assuming the scalar field has a harmonic time dependence $\phi(r,t) = \phi_0(r) e^{-i \omega t}$, where $r$ is the radial spherical coordinate. This condition implies that the stress energy tensor and the geometry of the space-time are time-independent. The construction of BS solutions assumes the metric in normal coordinates $ds^2=-\alpha(r)^2dt^2 + a(r)^2dr^2 + r^2 d\Omega^2$ and the Einstein-Klein-Gordon system reduces to the following set of ODEs

\begin{eqnarray}
\frac{\partial_r a}{a} &=& \frac{1-a^2}{2r} +\nonumber\\
	&&\frac{1}{4}\kappa_0 r
	\left[\omega^2 \phi_{0}^{2}\frac{a^2}{\alpha^2}
	+(\partial_r \phi_0)^{2} + 
	a^2 \phi_{0}^{2} (m^2 
	+ \frac{1}{2}\lambda \phi_{0}^{2})
	\right],\nonumber\\
\frac{\partial_r \alpha}{\alpha} &=&
	\frac{a^2-1}{r} + 
	\frac{\partial_r a}{a} -
	\frac{1}{2}\kappa_0 r a^2\phi_{0}^{2}(m^2 
	+\frac{1}{2}\lambda\phi_{0}^{2}),\nonumber\\
\partial_{rr}\phi_0  &+& \partial_r \phi_0  \left( \frac{2}{r} + 
	\frac{\partial_r \alpha}{\alpha} - \frac{\partial_r a}{a}\right) 
	+ \omega^2 \phi_0 \frac{a^2}{\alpha^2} \nonumber\\
	&-& a^2 (m^2 + \lambda \phi_{0}^{2}) \phi_0 
	=0,\label{sphericalekgc-sc}
\end{eqnarray}

\noindent which is solved under the conditions $a(0)=1$, $\phi_0(0)$ finite and $\partial_r \phi_0(0)=0$ in order to guarantee regularity and spatial flatness at the origin, and $\phi_0(\infty)=0$ in order to ensure asymptotic flatness at infinity as described in \cite{RuffiniBonazzola1969,Balakrishna1998,scott,Guzman2009}; these conditions together with  
(\ref{sphericalekgc-sc}) define an eigenvalue problem for $\omega$. For every central value of 
$\phi_0$ there is a unique $\omega$ that satisfies the boundary conditions. Rescaling the variables 
$\phi_0 := \sqrt{\frac{\kappa_0}{2}} \phi_0$, 
$r := mr$, 
$t := \omega t$, 
$\alpha := \frac{m}{\omega}\alpha$ and 
$\Lambda := \frac{2\lambda}{\kappa_0 m^2}$ the system above is written independently of $m$ and $\omega$:

\begin{eqnarray}
\frac{\partial_r a}{a} &=& \frac{1-a^2}{2r} +\nonumber\\
	&&\frac{1}{2} r
	\left[\phi_{0}^{2}\frac{a^2}{\alpha^2}
	+(\partial_r \phi_0)^{2} + 
	a^2 (\phi_{0}^{2}
	+ \frac{1}{2}\Lambda \phi_{0}^{4})
	\right],\nonumber\\
\frac{\partial_r \alpha}{\alpha} &=&
	\frac{a^2-1}{r} + 
	\frac{\partial_r a}{a} -
	r a^2\phi_{0}^{2}(1 
	+\frac{1}{2}\Lambda\phi_{0}^{2}),\nonumber\\
\partial_{rr}\phi_0  &+& \partial_r \phi_0  
	\left( 
		\frac{2}{r} + 
		\frac{\partial_r \alpha}{\alpha} - \frac{\partial_r a}{a}
	\right) 
	+ \phi_0 \frac{a^2}{\alpha^2} \nonumber\\
	&-& a^2 (1 + \Lambda \phi_{0}^{2}) \phi_0 
	=0.
\label{sphericalekgc-sc-rescaled}
\end{eqnarray}

\noindent Now the eigenvalue $\omega$ is included into the central value of the lapse $\alpha$ due to the rescaling. We solve (\ref{sphericalekgc-sc-rescaled}) using finite differences with an adaptive step-size fourth order Runge-Kutta and a shooting routine that bisects the central value of $\alpha$ and therefore calculates $\omega$.


\begin{figure}[htp]
\includegraphics[width=8cm]{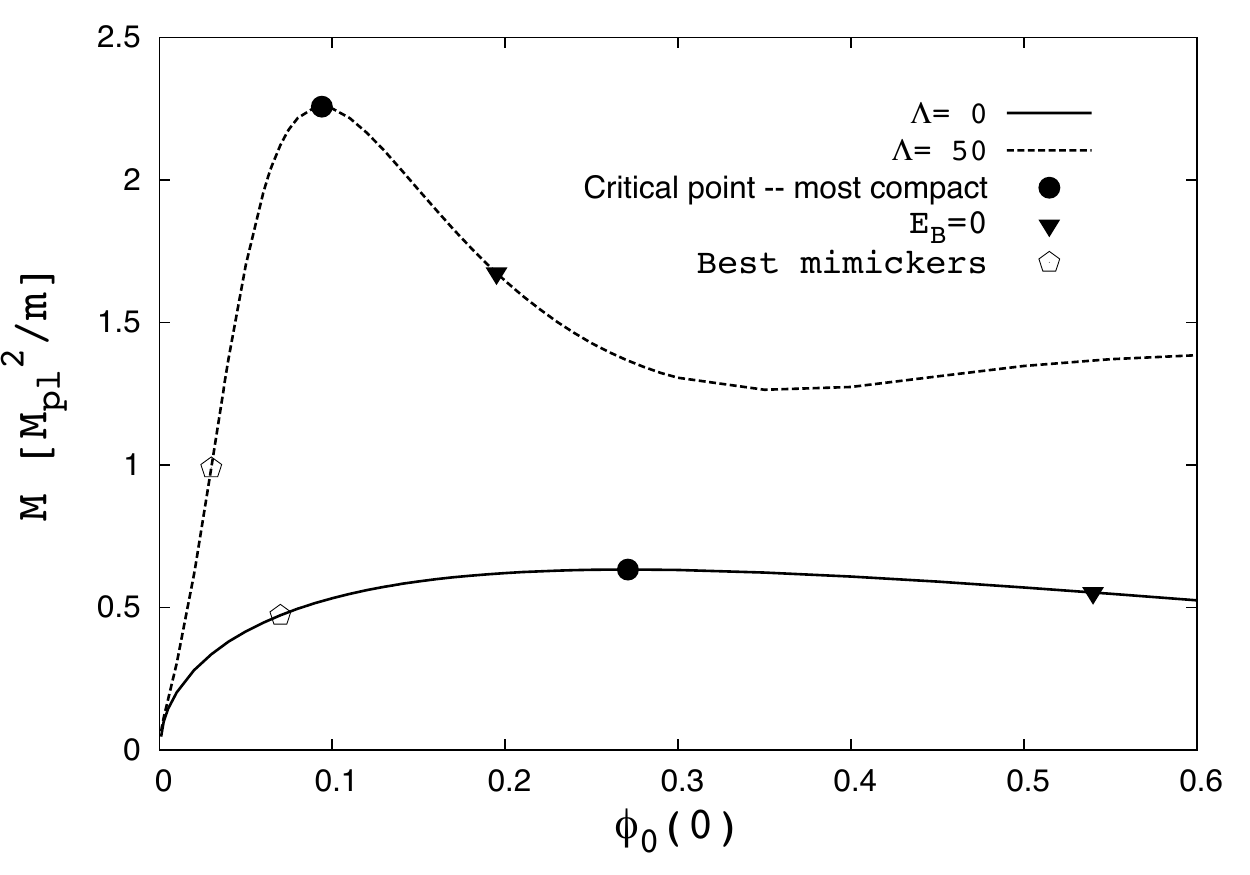}
\caption{\label{fig:equilibrium} Equilibrium configurations for the two values of $\Lambda$ used in this paper in an $M$ vs $\phi_0(0)$ diagram. Each point in the curves corresponds to a solution of the eigenvalue problem and represents a BS configuration. The filled circles indicate the critical solution that separates the stable from the unstable branch and is also the most compact of the stable configurations. Those configurations to the left of the maxima represent stable configurations. For completeness we also present an extended view of the unstable branch. The inverted triangles indicate the point at which the binding energy is zero. Those configurations between the filled circles and the inverted triangles (along each sequence) collapse into black holes as a response to a perturbation. Configurations to the right of the inverted triangles disperse away. The pentagons correspond to the best black hole mimickers for $\Lambda=0$ and $\Lambda=50$ when a stationary accretion disk model is used to generate a luminosity power spectrum.}
\end{figure}

The solutions of (\ref{sphericalekgc-sc-rescaled}) define sequences of equilibrium configurations, one for each value of $\Lambda$. In Fig. \ref{fig:equilibrium} we show the sequences of equilibrium configurations for $\Lambda=0$ and $\Lambda=50$ in order to consider a case with non-zero self-interaction. Each point in the curves corresponds to a BS solution for a given central field value. In each of the curves three important points for each value of $\Lambda$ are marked: 

\begin{itemize}
\item[i)] the critical point -marked with a filled circle- indicating the threshold between the stable and unstable branches of each sequence, explicitly, configurations to the left of this point are stable and those to the right are unstable \cite{scott,SeidelSuen1990,Balakrishna1998,Guzman2004}.

\item[ii)] the point at which the binding energy $E_B = M-Nm = 0$ marked with an inverted filled triangle, where $N=\int 
j^0 d^3 x = \int \frac{i}{2}\sqrt{-g}g^{\mu\nu}[\phi^{*}\partial_{\nu}\phi - \phi \partial_{\nu}\phi^{*}]d^3 x$ is the number of particles; that is, the conserved quantity due to the invariance under the global $U(1)$
group of the Lagrangian density. $M = (1-1/a^2)r/2$ evaluated at the outermost point of the numerical domain is the Misner-Sharp mass; the configurations between the instability 
threshold and the zero binding energy point have negative binding energy
($E_B<0$) and collapse into black holes whereas those to the right 
have positive binding energy and disperse away \cite{Guzman2004,Guzman2009}. The astrophysically relevant branch is the stable branch which contains configurations that are not expected to collapse or disperse away under perturbations.

\item[iii)] The points marked with a pentagon correspond to two Boson Stars, one in the curve for $\Lambda=0$ another for $\Lambda=50$, that best mimic a black hole of the same mass using a stationary accretion disk model \cite{Chucho2009}.

\end{itemize}

\subsection{Selection of BSs}

We want to compare some properties of the gas distribution during the interaction with a BS configuration. In particular it is interesting to know how the self-interaction $\Lambda$ and the compactness of the BS affect the distribution of the gas. In order to investigate the effects of $\Lambda$ one can choose any stable configuration for the two values of $\Lambda=0,50$, as long as they are stable, then the differences in the gas configuration will be due to the value of $\Lambda$. Instead of choosing any two configurations we choose two that have been used as Black Hole mimickers in the context of stationary accretion disks \cite{Chucho2009}. The configuration for $\Lambda=0$ and that for $\Lambda=50$ are marked with pentagons in the Figure. These configurations have the following parameters:

\begin{itemize}
\item[1.] Best mimicker with $\Lambda=0$  (BSBM$\Lambda 0$), with parameters $\phi_0(0)=0.07$ and $M=0.47297[M^{2}_{pl}/m]$.
\item[2.] Best mimicker with $\Lambda=50$ (BSBM$\Lambda 50$), with parameters $\phi_0(0)=0.03$ and $M=0.9898[M^{2}_{pl}/m]$.
\end{itemize}

On the other hand, we study the effects of compactness by choosing the most compact BSs for the case $\Lambda=0$. The most compact configuration is expected to be more similar to a black hole in terms of the gravitational potential that generates. One needs to keep in mind that the most compact configuration of the stable branch is also at the onset of instability, and therefore any perturbation in full General Relativity would collapse it into black holes, nevertheless, being the most compact will help to understand the effect of compactness. The configuration we use is pointed out in Fig. \ref{fig:equilibrium} with a filled circle along the family $\Lambda=0$, and its parameters are:

\begin{itemize}

\item[3.] Most compact with $\Lambda=0$ (MCBS$\Lambda 0$), with parameters $\phi_0(0)=0.271$ and $M=0.633[M^{2}_{pl}/m]$.

\end{itemize}

These are the three BS configurations we focus on along this paper.

\subsection{The wind}

First of all, it is worth mentioning that these BSs are spherically symmetric and therefore non-rotating. Thus, in order to make an appropriate comparison we consider non-rotating BHs. Therefore the orientation of the wind with respect to the BS or the BH is not important. The effects of the BS on the wind have to be compared with the effects produced by a Schwarzschild black hole of the same mass on the properties of the wind variables

We simulate the evolution of the wind by solving the relativistic Euler equations on a curved space-time corresponding to a BS or a BH. The wind propagates initially along a given direction with a given asymptotic velocity $v_{\infty}^2=v^iv_i$. We assume the gas initially has a spatially constant rest mass density. It also obeys a gamma-law equation of state $p=(\Gamma-1)\rho \epsilon$, that we use to calculate the asymptotic speed of sound $c_{s \infty}$. Once we define the value of $c_{s\infty}$ and assume the density to be initially a constant $\rho = \rho_{\mathrm{ini}}$, the pressure can be written as $p_{\mathrm{ini}} = c_{\mathrm{s} \infty}^2 \rho_{\mathrm{ini}}/(\Gamma - c_{\mathrm{s} \infty}^2 \Gamma_1)$, where $\Gamma_1=\Gamma/(\Gamma -1 )$. In order to avoid negative and zero values of the pressure, the condition $c_{\mathrm{s}\infty} < \sqrt{\Gamma - 1}$ has to be satisfied. Finally, with this value for $p_{\mathrm{ini}}$, the initial internal specific energy is reconstructed using the equation of state. In this paper we set $\Gamma=4/3$.

We also reduce the analysis to two regimes of wind speed, a fast and a slow wind. This is also equivalent to choose different relative scales between the object radius and the accretion radius (see e.g. \cite{Mycol2015}). The specific properties of the fast and slow regimes are as follows

\begin{itemize}
\item[] Fast wind. $\sqrt{v_{\infty}^2}=0.5c$, $r_{acc}=3.63M$, $\rho_0 = 1\times 10^{-6}M^{-2}$.
\item[] Slow wind. $\sqrt{v_{\infty}^2}=0.25c$, $r_{acc}=47.297M$ $\rho_0 = 1\times 10^{-6}M^{-2}$.
\end{itemize}

\noindent where the accretion radius is $r_{acc}= 2M/(v_{\infty}^2+c_{s\infty}^2)$ and $M$ is the mass of the accretor, in our case a BS or a BH. The scale of $r_{acc}$ fixes also the size of the accretor and determines the size of the numerical domain to be used. It is important that a sphere of radius $r_{acc}$ fits within the numerical domain. Notice that $v_{\infty}$ is in units of $c$, $r_{acc}$ in units of the accretor mass $M$ and the density in units of $M^{-2}$. More details related to the relation between initial conditions and the size of an appropriate domain, as well as more general wind configurations can be found in \cite{Mycol2015}.

In all the cases the space-time of BSs and BHs is described in cartesian coordinates and the domain is a cube centered at the origin and $(x,y,z) \in [x_{min},x_{max}]\times[y_{min},y_{max}]\times[z_{min},z_{max}]$. We choose the wind to move along the $x$ direction injected with the velocity and properties described above from the face $x=x_{min}$. The half of the domain $x <0$ defines the upstream region whereas $x>0$ is the downstream region of the fluid.

\subsection{Numerical Setup}

We solve the Relativistic-Euler equations using the Cactus Einstein Toolkit (ETK)  \cite{ETK2012,ETK} which provides the necessary computational tools to evolve this relativistic fluid in full 3D on an arbitrary space-time background. Specifically we use the GRHydro Thorn \cite{baiotti2005}. This application uses  high resolution shock capturing methods to solve these equations. We specifically use the HLLE flux formula and the minmod linear reconstructor. For the integration in time we use the method of lines with a fourth order Runge-Kutta integrator. We use our own Boundary Conditions module that allows the use of different conditions on different faces of the cubic domain. In the boundary of the upstream region at the face $x=x_{min}$, we inject the wind with velocity and density mentioned above toward the BS or the BH. In the other five faces of the cubic domain we implement out-flux boundary conditions. The numerical domain uses three refinement level resolutions $\Delta_1 = 2\Delta_2 = 4\Delta_3 = 0.25M$, defined and controlled in the ETK with the Carpet driver \cite{carpet}.

The space-time of the BS constructed in subsection \ref{subsec:BSs} requires some preprocessing before starting an evolution. The solution is written in spherical coordinates and therefore it has to be  interpolated from a numerical grid in spherical coordinates to a 3D cartesian grid defined within the ETK. After that, the metric functions are converted from spherical to cartesian coordinates. Notice that since we are assuming the space-time is fixed, or equivalently the wind does not distort the space-time, the scalar field is not required to evolve or play any role during the evolution because the scalar field does not interact with the wind. Furthermore, since the bosons interact only gravitationally with the wind, only the geometry of the BS space-time is required. In the end, physically the wind will come from an asymptotic region, approach the potential well of the BS and change its streamlines and density. A similar effect happens with gravitational lensing by BSs, that is, there is not a photon sphere and light is only deflected by the BS \cite{RuedaGuzman2009}.

In the case of the Black Hole part of the gas enters the black hole's horizon. Again, we assume the space-time is fixed and therefore the BH does not grow due to the accretion. We describe the space-time of the Schwarzchild Black Hole in Eddington-Finkelstein coordinates so that we can use the excision inside the black hole horizon, which is already implemented in the ETK. The excision method consists in removing a piece of the numerical domain inside the black hole's apparent horizon in order to avoid the black hole singularity. This removal defines a numerical inner boundary. In horizon penetrating coordinates the light cones at this boundary are open, pointing toward the singularity and therefore both, geometrical and material quantities are regular there. The advantage at this boundary is that the coordinates themselves help to pull the material particles, in our case the fluid elements inside the horizon are absorbed by this boundary without the need of boundary conditions as described in \citep{hawke2005} for the accretion of matter into black holes using the ETK. This method is needed only for the BH space-time because a BS does not have horizon nor singularity.

\section{Results}
\label{sec:results}

\subsection{Case $\Lambda=0$.}

Knowing that BSs have no surface, nor horizon and ordinary matter interacts with the bosons only through gravity, it is expected that an incoming wind will find a slightly bent space-time and suffer a reconfigurations of the streamlines and a possible accumulation of gas and then escape to infinity, because there is no accretion expected to happen during the process. 

\begin{figure*}[htp]
\includegraphics[width=8.5cm]{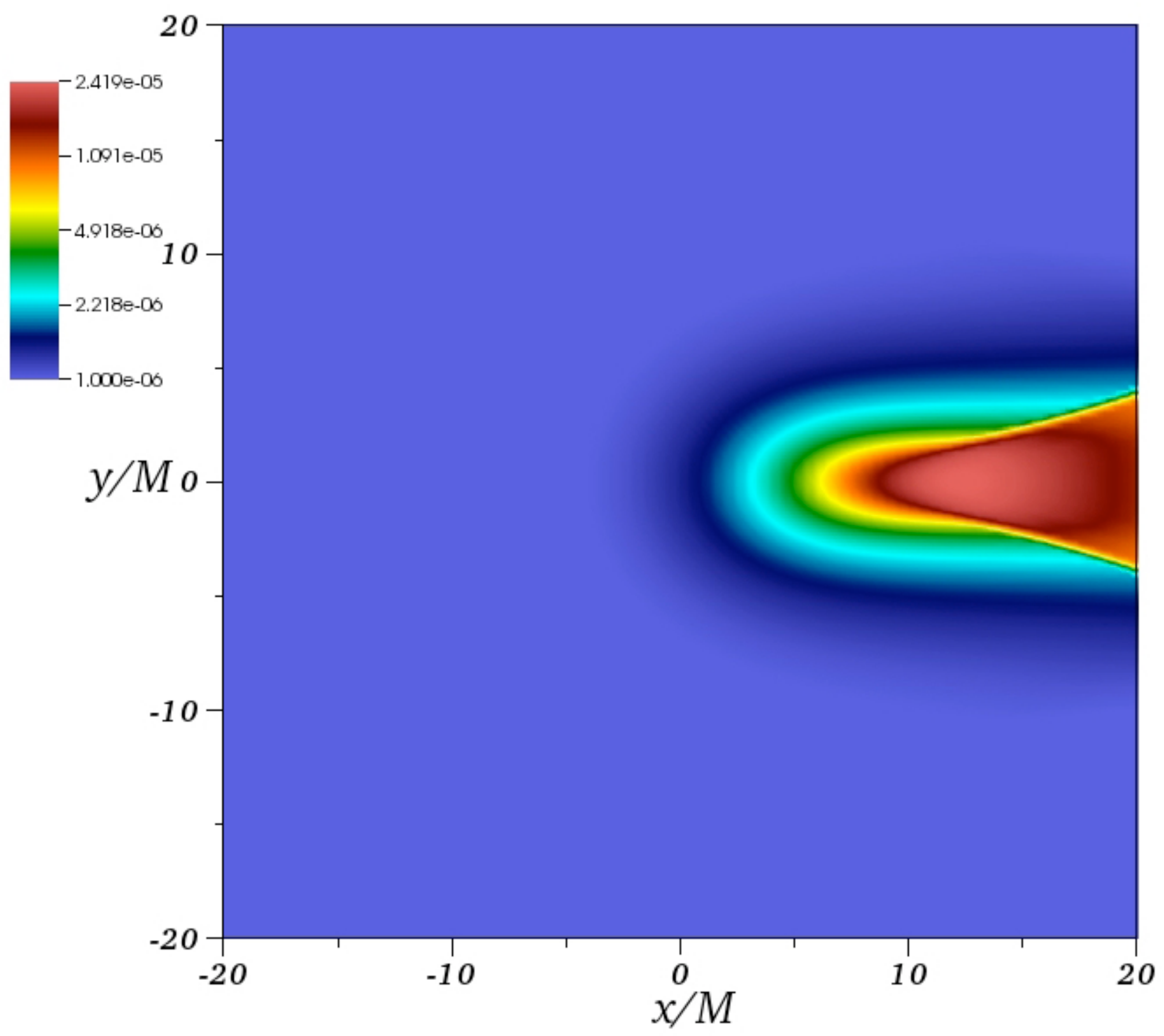}
\includegraphics[width=8.5cm]{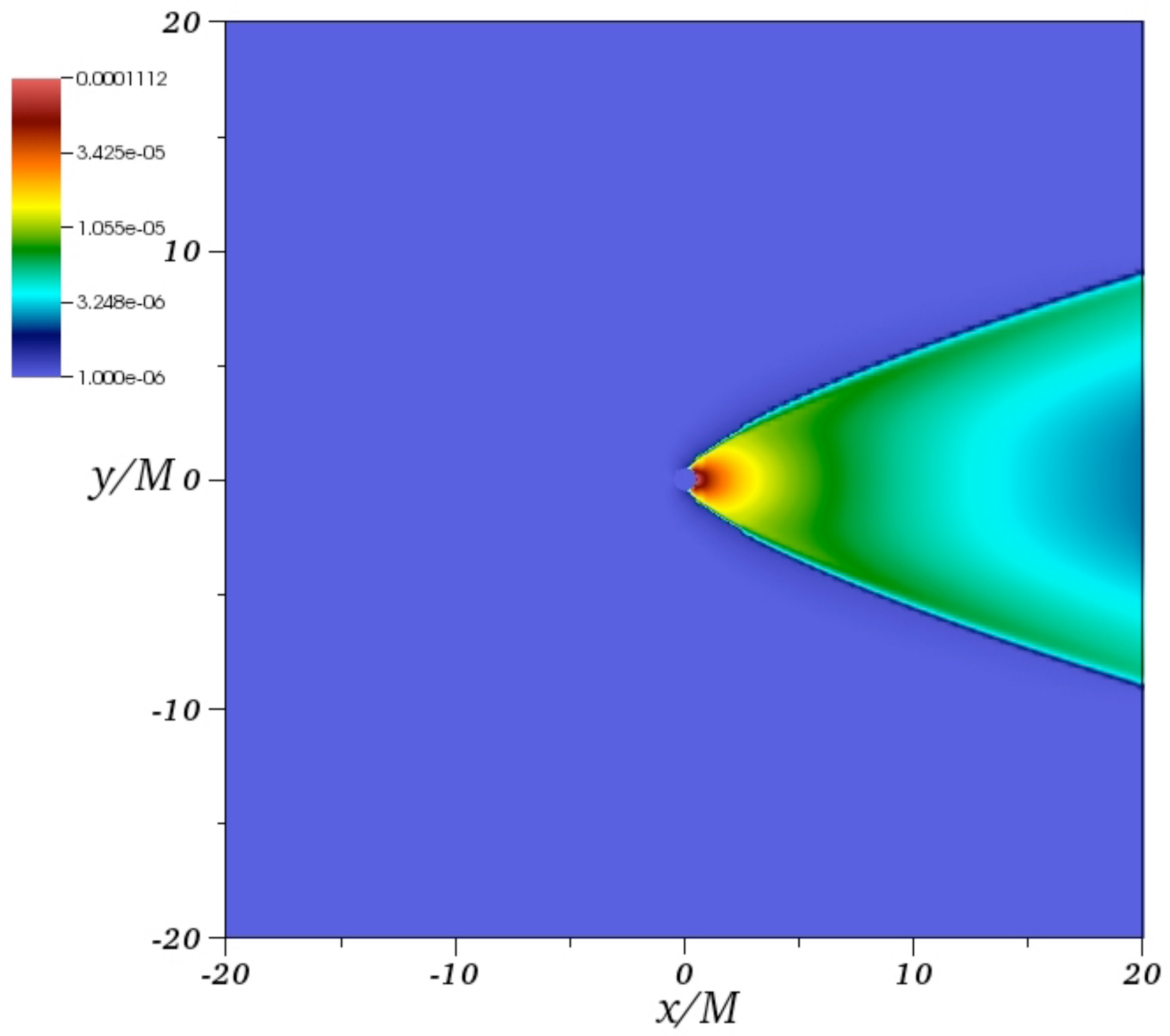}
\includegraphics[width=8.5cm]{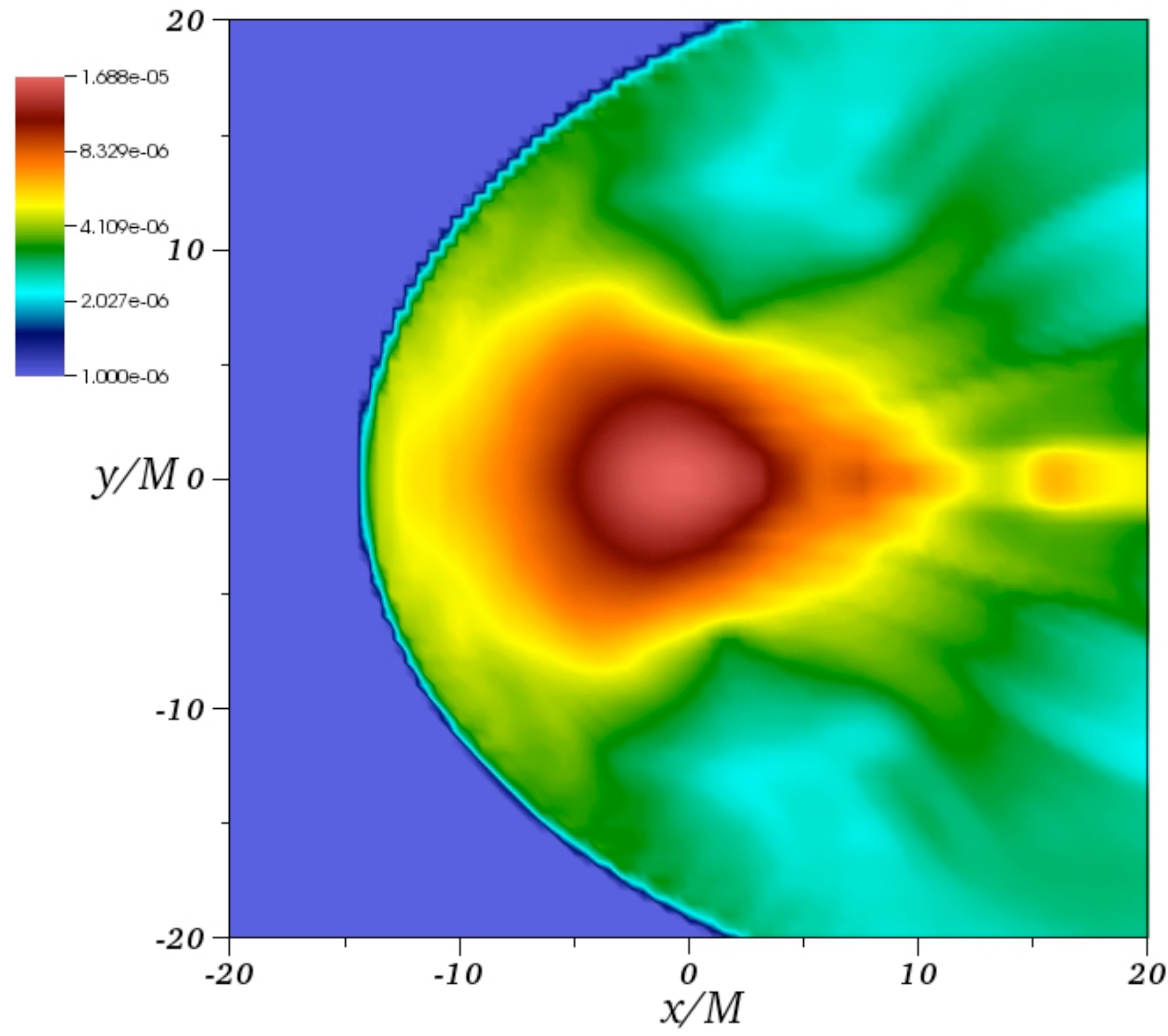}
\includegraphics[width=8.5cm]{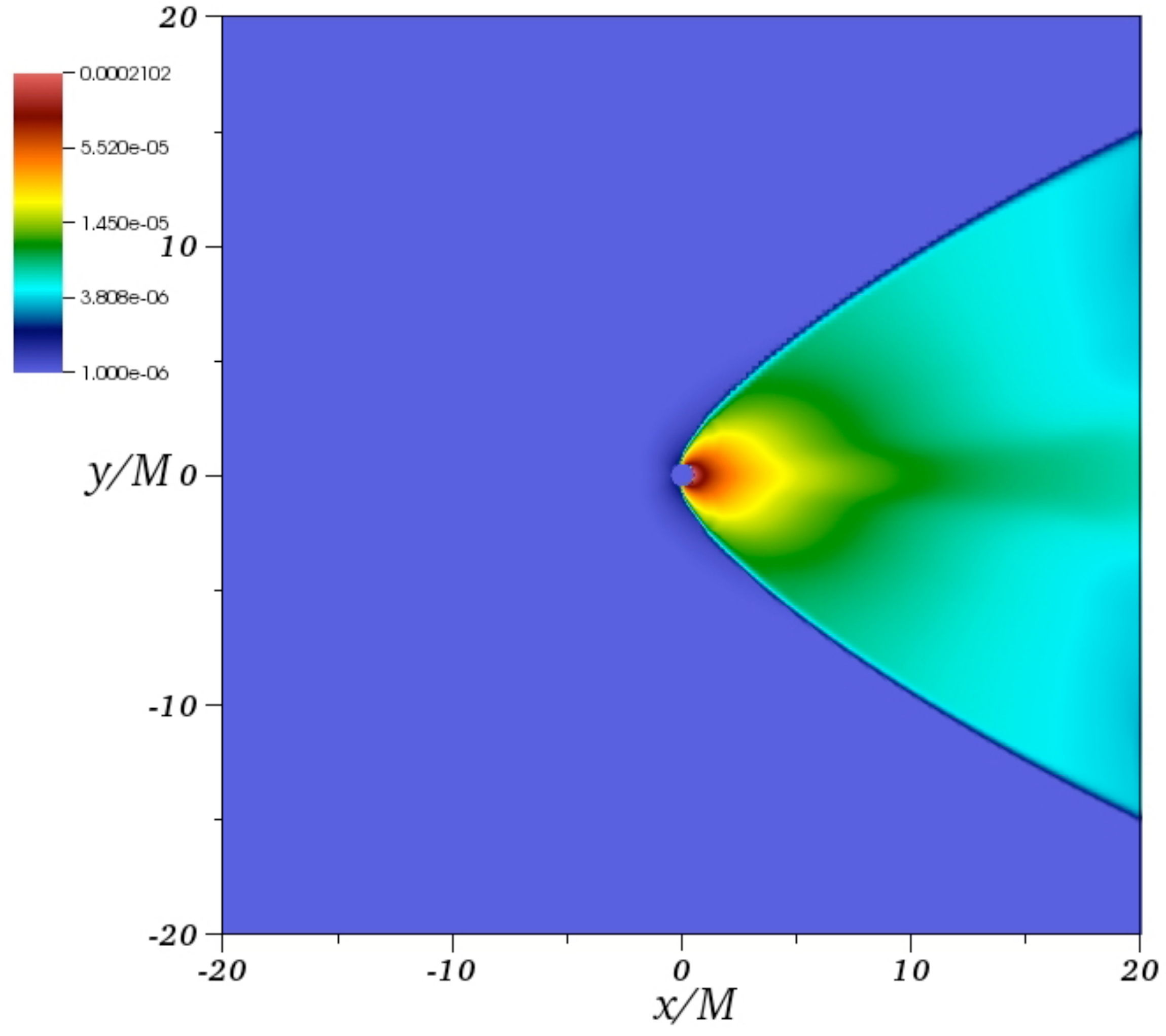}
\caption{\label{fig:BS1density} Density of the gas moving along the $x$ direction from left to right for the Best Mimicker without self-interaction BSBM$\Lambda 0$. In the top frame we show the fast case corresponding to an asymptotic velocity $0.5c$, whereas in the bottom panel we show a slower wind with velocity $0.25c$. The BSs are located at the center of coordinates, where the maximum density of the gas is accumulated. For comparison we show on the right the shock cone formed when instead of a Boson Star there is a Schwarzschild Black hole with the same mass. The numerical domain is different for the different wind velocities considering a sphere of radius $r_{acc}= 2M/(v_{\infty}^2+c_{s\infty}^2)$ needs to be contained within the numerical domain. The circle inside the black hole is the excision region on the plane $z=0$.}
\end{figure*}

What the simulations show is that after an initial transient, like in the accretion onto a BH, the configuration approaches a stationary regime with a high gas density elongated shape, formed in the downstream zone of the Boson Star. We show the gas density in Fig. \ref{fig:BS1density} for the configuration BSBM$\Lambda$0 and a black hole with the same mass, using the two wind speeds. 

The morphology indicates that the high density region is different from the shock cone formed behind the BH. In the case of the {\sl fast} wind there is an important density gradient, and the density of this structure is about five times smaller than the shock cone behind the BH. In the {\sl slow} case the density is one order of magnitude smaller.  This means that when the wind is slower, the differences in the high density of the gas around the BS and on the BH are more important.

Also the morphology for the slow wind has more structure than in the fast wind case. It calls the attention that the density on the bottom-left frame of Fig. \ref{fig:BS1density} is actually stationary. We remind the reader that on the left boundary of this image in the upstream zone we maintain a constant wind entering the numerical domain, whereas on all the other boundaries we use out-flux boundary conditions. 

\begin{figure*}[htp]
\includegraphics[width=8.5cm]{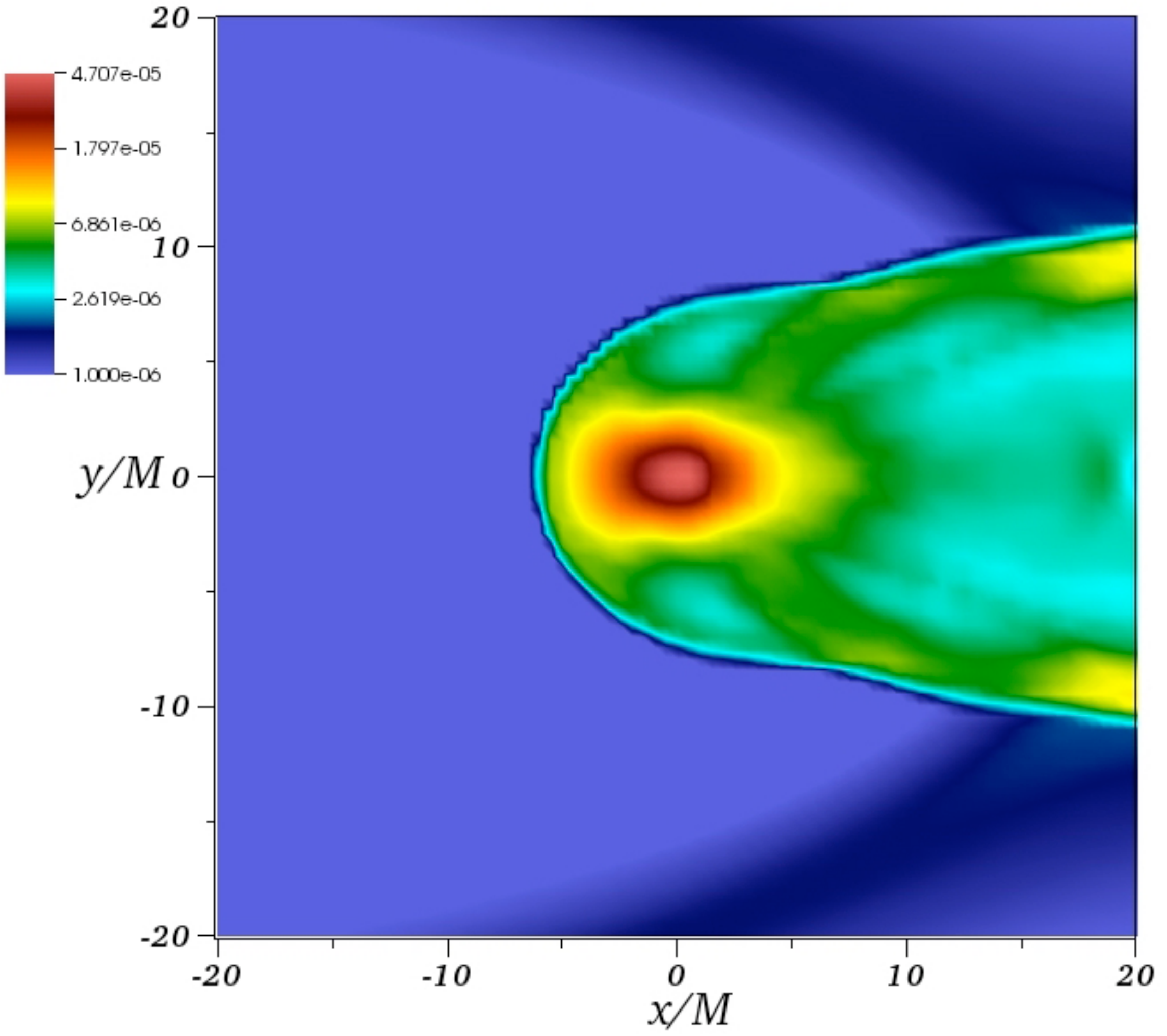}
\includegraphics[width=8.5cm]{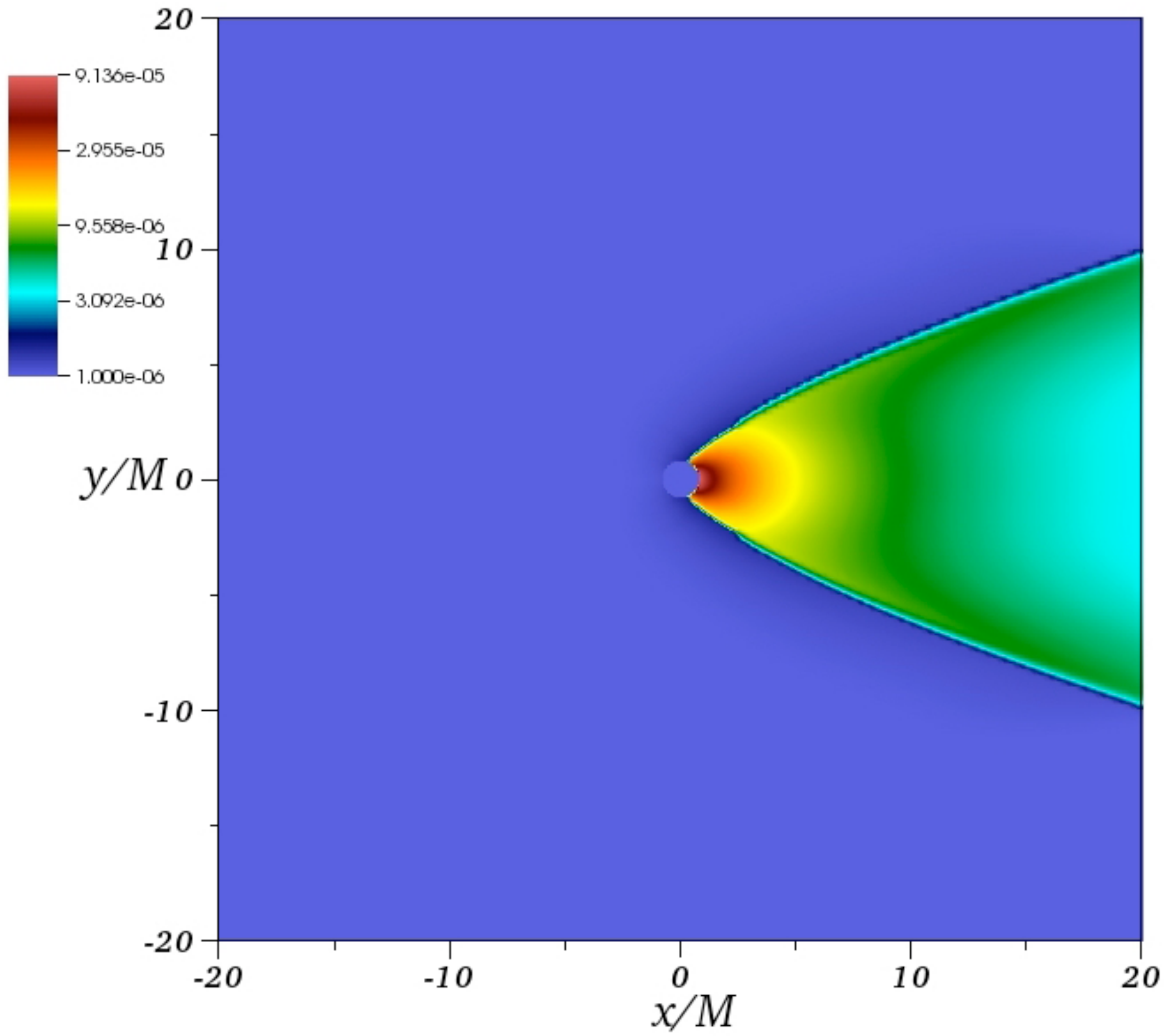}
\caption{\label{fig:BS3density} Density of the gas moving from left to right with velocity $0.5c$. On the  left the case of the most compact BS MCBS$\Lambda 0$, and on the right the case of the wind on the space-time of a BH of the same mass.}
\end{figure*}

In order to know the effects of the compactness of the BS, in Fig. \ref{fig:BS3density} we show the results for the most compact BS with $\Lambda=0$ and velocity 0.5$c$, together with the shock cone due to a BH of the same mass. The density profile again approaches a stationary regime. The morphology of the high density structure around the BS is different from that of BSBM$\Lambda 0$ in the top-left of Fig. \ref{fig:BS1density}. The highest density of the gas is located at the center of the BS and is nearly half of the maximum density of the gas of the shock cone on the BH space-time. This means that the more compact the BS is, the more similar the highest density is to that of the shock cone in a BH space-time.

\subsection{Case $\Lambda=50$.}

The case with self-interaction is well known for allowing BS to be more massive. We thus show the results for the best black hole mimicker with $\Lambda=50$ according to \cite{Chucho2009}. The morphology of high density elongated structure is nearly the same as that without self-interaction. The highest density of the gas is about one third the highest density of the shock cone on a BH of the same mass for the fast wind, whereas it is nearly one fifth in the case of the slow wind.  This means that when considering self-interaction the density of the gas is more similar to that of a BH than when using $\Lambda=0$. A final observation on the morphology is that for the slow wind, a structure similar to a bow shock seems to be evolving on the upstream zone, nevertheless this is actually not moving and remains stationary.

\begin{figure*}[htp]
\includegraphics[width=8.5cm]{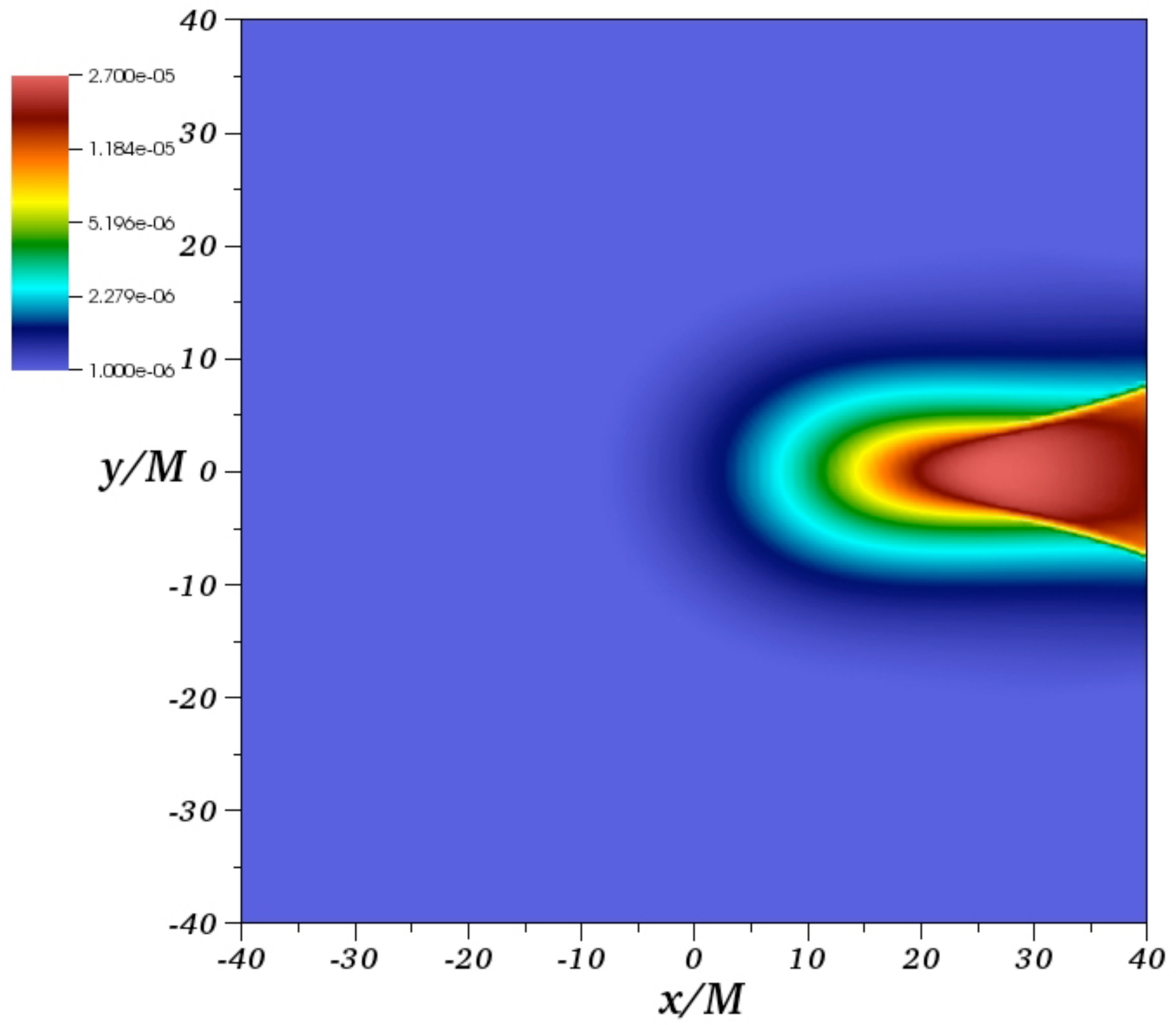}
\includegraphics[width=8.5cm]{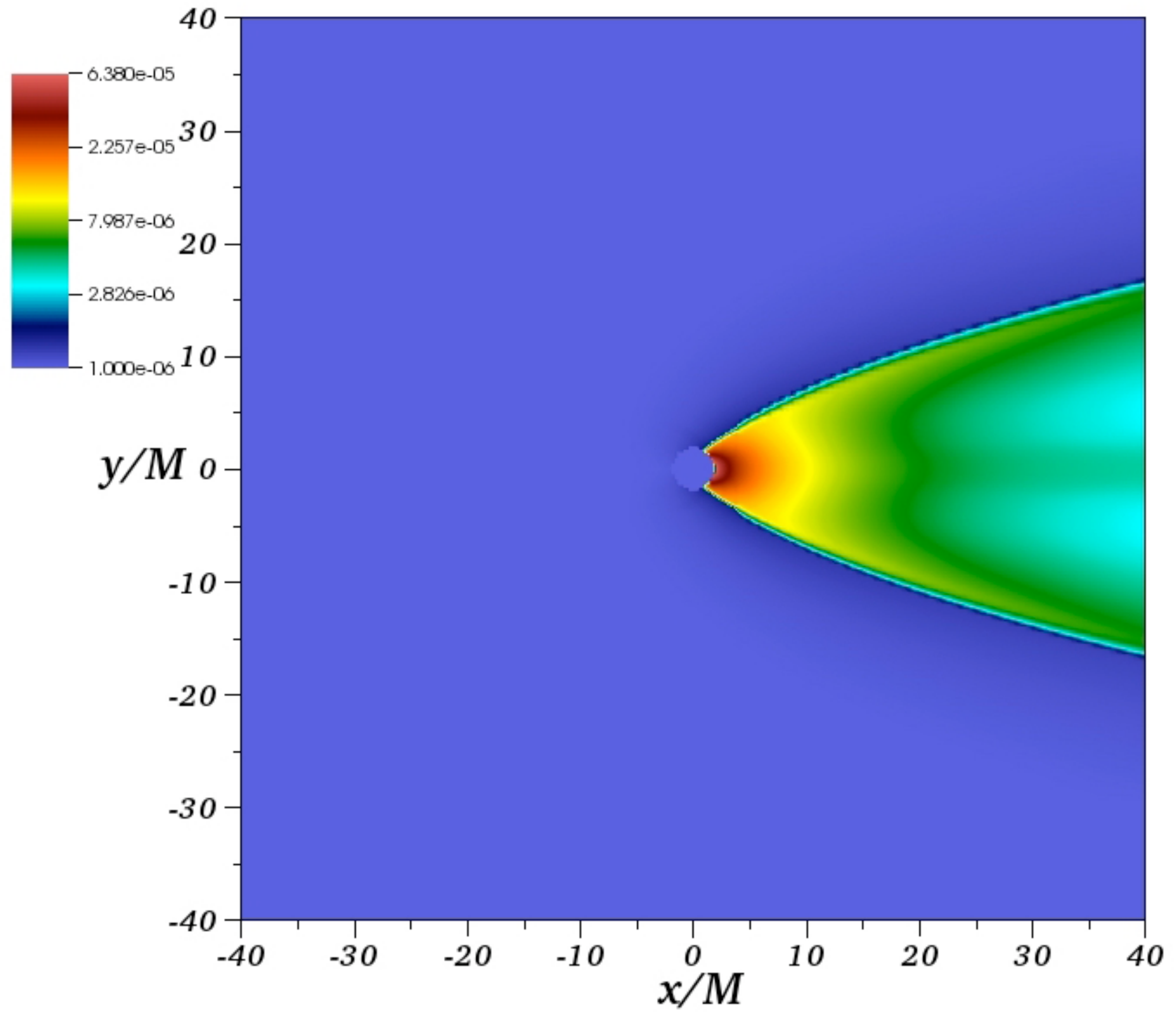}
\includegraphics[width=8.5cm]{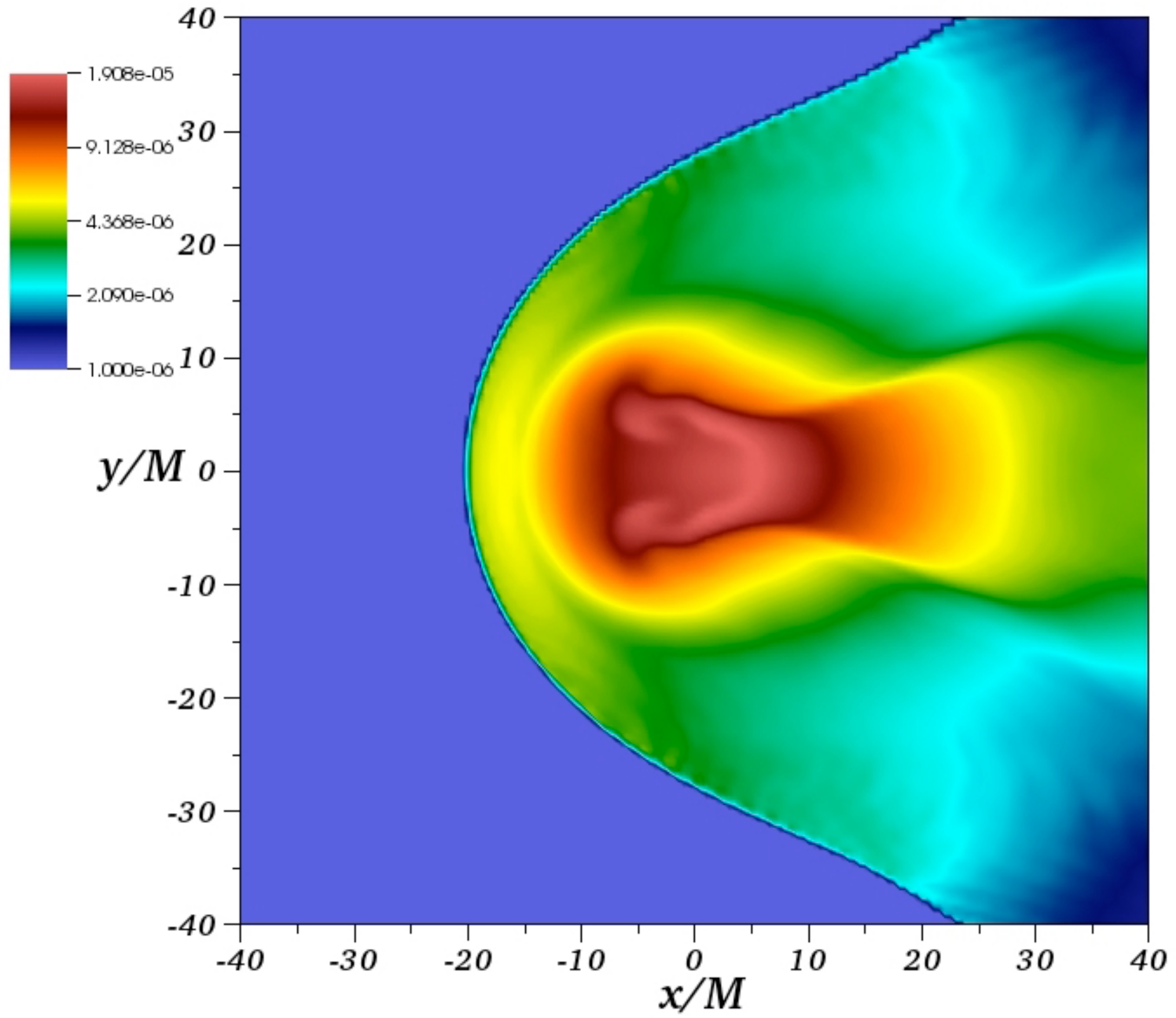}
\includegraphics[width=8.5cm]{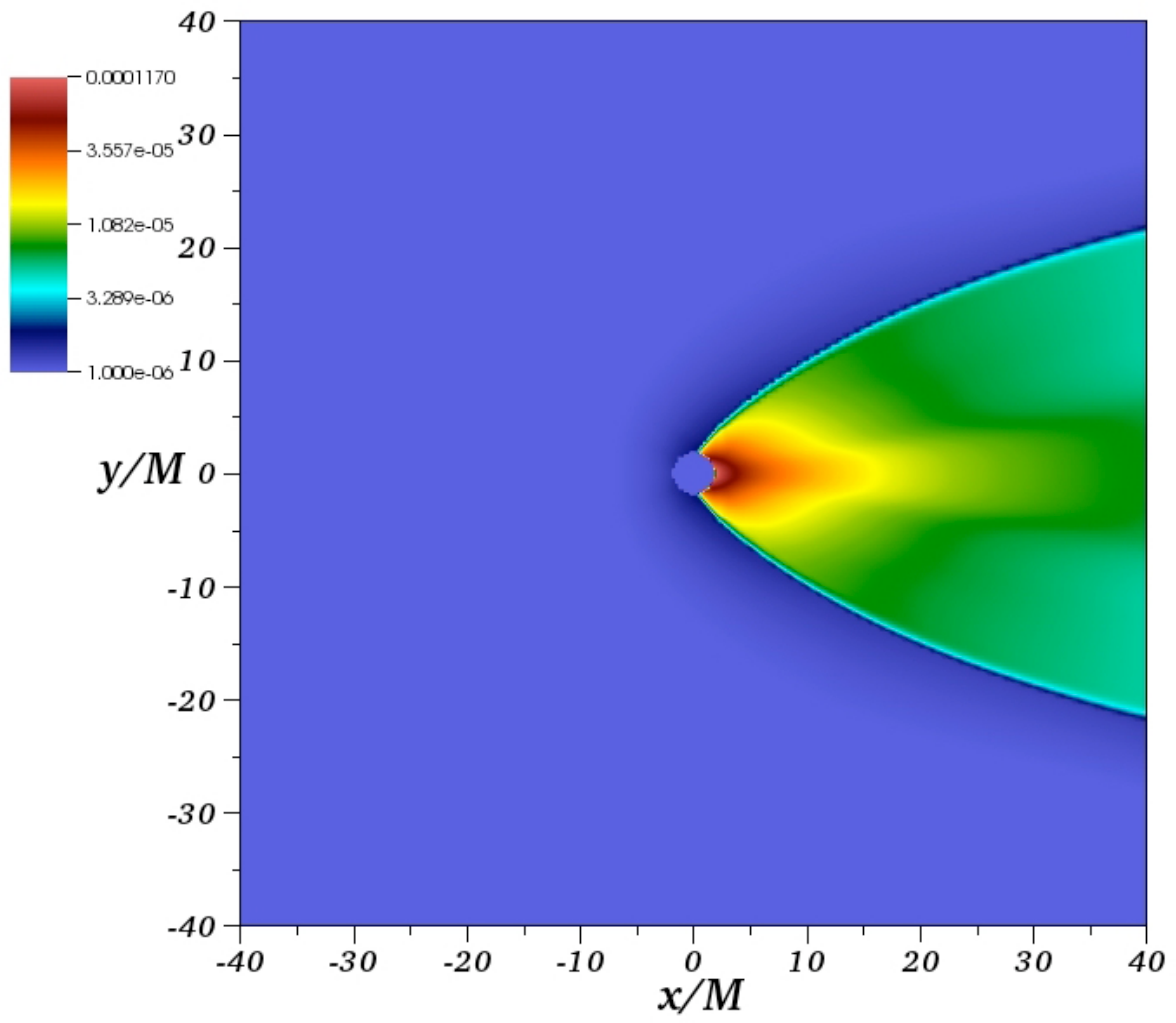}
\caption{\label{fig:BS2density} Density of the gas moving from left to right for the Best Mimicker BSBM$\Lambda 50$. As in the case without self-interaction, on the left we show a snapshot of the density once the high density structure has stabilized and on the right we show the shock-cone formed when the accretor is a black hole of the same mass. The velocities are again $0.5c$ (top) and $0.25c$ (bottom).}
\end{figure*}

\section{Conclusions}
\label{sec:conclusions}

We studied the Bondi-Hoyle accretion process of a wind onto a Boson Star space-time and compared it with the process when the wind is accreted by a Black Hole of the same mass. For this we simulated the evolution of a wind consisting of an ideal gas with fast and slow supersonic velocities.

In order to learn about the impact of self-interaction we studied the process on two typical BS configurations with $\Lambda=0$ (BSBM$\Lambda 0$) and $\Lambda=50$ (BSBM$\Lambda 50$). In the two cases, for the fast and slow winds, the density reaches a stationary configuration with a high density elongated structure around the Boson Star. 

For the fast (slow) wind case we found that the density is smaller on a BS than when the accretion happens on a BH with the same mass by a factor of five (twelve) for $\Lambda=0$, whereas when the BS has a high self-interaction $\Lambda=50$ the density of the gas in the stationary regime is smaller than in the BH by a factor of three (six). This indicates on the one hand, that a configuration with higher self-interaction produces a gas density more similar in magnitude to that of the shock cone on a BH. On the other hand these results show that the faster the wind, the more similar the density of the structure.

We also wanted to explore the influence of the compactness of the BS. For that we use the configuration (BSBM$\Lambda 0$) and compare it with the most compact BS for $\Lambda=0$  (MCBS$\Lambda 0$) for the fast wind. For the most compact configuration we found that the peak density of the gas is smaller than with the BH only by a factor of two. Thus we found that when the BS is more compact, the density of the stationary gas configuration is more similar to that of the shock cone formed in the downstream zone of a BH.

Collecting these results together, we found that the gas density in the Bondi-Hoyle process on a BS space-time is more similar to that of a shock cone around a BH space-time when the wind is fast, the BS has positive self-interaction and when it is more compact.

These results are peculiar. BSs have no event horizon, nor mechanical surface, they do not accrete in the sense that they trap the gas, and nevertheless they are able to concentrate the gas density in a small region with high density. We have shown here some morphological differences in the gas configuration, namely, the shape of the high density structure and the magnitude of the density itself. However, the gas density is comparable for BSs and BHs. In order to provide a quantitative comparison in observational grounds we are planning to perform the analysis using Relativistic Radiation Hydrodynamics \cite{Pancho2016} in order to estimate luminosity curves of the process.


\section*{Acknowledgments}

This research is partly supported by grants CIC-UMSNH-4.9, CONACyT 258726 (Fondo Sectorial de Investigaci\'on para la Educaci\'on). The simulations were carried out in the IFM Draco cluster funded by  CONACyT 106466.


\end{document}